\documentclass[preprint2]{aastex}

\slugcomment{Not to appear in Nonlearned J., 45.}

\shorttitle{The nature and nurture of bars and disks}
\shortauthors{M\'endez-Abreu et al.}

\begin{document}


\title{The nature and nurture of bars and disks}


\author{J. M\'endez-Abreu\altaffilmark{1,2}, R. S\'anchez-Janssen\altaffilmark{3}, J. A. L. Aguerri\altaffilmark{1,2}, E. M. Corsini\altaffilmark{4,5}, and
S. Zarattini\altaffilmark{1,2,4}}

\altaffiltext{1}{Instituto de Astrof\'isica de Canarias, Calle V\'ia L\'actea s/n, E-38200 La Laguna, Tenerife, Spain}
\altaffiltext{2}{Departamento de Astrof\'isica, Universidad de La Laguna,  E-38205 La Laguna, Tenerife, Spain}
\altaffiltext{3}{European Southern Observatory, Alonso de Cordova 3107, Vitacura, Santiago, Chile}
\altaffiltext{4}{Dipartimento di Fisica e Astronomia `G. Galilei', Universit\`a di Padova, vicolo dell'Osservatorio 3, I-35122 Padova, Italy}
\altaffiltext{5}{INAF--Osservatorio Astronomico di Padova, vicolo dell'Os\-ser\-vatorio 5, I-35122 Padova, Italy}


\begin{abstract}
The effects  that interactions  produce on galaxy  disks and  how they
modify the  subsequent formation of  bars need to be  distinguished to
fully understand the relationship between bars and environment.
To this aim we derive the bar fraction in three different environments
ranging  from  the field  to  Virgo  and  Coma clusters,  covering  an
unprecedentedly large range  of galaxy luminosities (or, equivalently,
stellar masses). We confirm  that the fraction of barred galaxies
  strongly  depends  on  galaxy  luminosity.  We also  show  that  the
  difference between  the bar fraction distributions as  a function of
  galaxy  luminosity (and  mass)  in  the field  and  Coma cluster  are
  statistically significant, with Virgo being an intermediate case.
The fraction  of barred  galaxies shows a  maximum of about  $50\%$ at
$M_r\,\simeq\,-20.5$  in  clusters, whereas  the  peak  is shifted  to
$M_r\,\simeq\,-19$ in the field.
We interpret this  result as a variation of  the effect of environment
on  bar formation depending  on galaxy  luminosity. We  speculate that
brighter disk galaxies are  stable enough against interactions to keep
their cold structure,  thus, the interactions are able  to trigger bar
formation. For fainter galaxies  the interactions become strong enough
to heat up the disks  inhibiting bar formation and even destroying the
disks.
Finally, we point  out that the controversy regarding  whether the bar
fraction  depends on  environment  could be  resolved  by taking  into
account the  different luminosity ranges probed by  the galaxy samples
studied so far.

\end{abstract}





\section{Introduction}
\label{sec:introduction}

It is  commonly accepted  that galaxy bars  spontaneously form  due to
instabilities  in dynamically cold  disks \citep{OstrikerPeebles1973}.
The growth rate of bars depends on the halo-to-disk mass ratio and the
velocity      dispersions      of      the     disk      and      halo
\citep{AthanassoulaSellwood1986} and  bars grow faster  in massive and
cold disks.  Moreover, environmental  processes can regulate  the life
cycle of bars  contributing both to their development  by forcing disk
instabilities   and   to    their   destruction   via   disk   heating
\citep[][]{Friedli1999}.

The role of environment in triggering the formation of bars has been a
matter of discussion  for a long time. A variety  of methods to detect
bars in  galaxy disks and measure  the local galaxy  density have been
adopted, but the  results on the correlation between  the bar fraction
and environment are controversial.
Since the early work by  \citet{Thompson1981} who claimed that the bar
fraction of  Coma galaxies increases  toward the core of  the cluster,
similar  results  were  found   for  the  Virgo  and  Fornax  clusters
\citep{Andersen1996,  Eskridge2000}   and  for  intermediate  redshift
clusters \citep{Barazza2009}.   Furthermore, some authors  measured an
increase  of  the  bar  fraction  in  galaxy  pairs  \citep{Kumai1986,
  Elmegreen1990,       Giuricin1993,      Varela2004}.       Recently,
\citet{Barway2011}  have  found that  fainter  lenticular galaxies  in
clusters show  a higher  bar fraction than  their counterparts  in the
field, and  \citet{Skibba2012}, using clustering  methods, have argued
that barred galaxies tend to populate high density environments.
In contrast, other  authors reported that environment does  not play a
major  role in bar  formation.  According  to \citet{vandenBergh2002},
and later to \citet{Aguerri2009}  and \citet{Li2009}, the bar fraction
strongly depends  on the  properties of the  host galaxies but  not on
their environment. Also \citet{Lee2012}  claimed that the bar fraction
does not  depend on the  environment when colour and  central velocity
dispersion      are      fixed.       \citet{MendezAbreu2010}      and
\citet{NairAbraham2010}  pointed  out that  galaxy  mass  is the  main
parameter driving the bar formation  and observed that bars are hosted
by galaxies  in a  tight range of  masses.  \citet{MartinezMuriel2011}
found that the bar population does not significantly depend neither on
group   mass  nor   on  the   distance  to   the   nearest  neighbour.
\citet{Giordano2011}   compared   two   carefully   selected   samples
representative   of    isolated   and   cluster    galaxies,   whereas
\citet{Marinova2012}  investigated  the  bar  fraction  in  lenticular
galaxies  across  different  environments  which span  two  orders  of
magnitude  in  galaxy  density.   Neither of  them  found  significant
differences.

Despite the  fact that  bars naturally appear  in simulations of
galaxy  formation once a  dynamically cold  and rotationally-supported
disk is in place \citep{Athanassoula2005},  the use of bars as tracers
of   disks    has   not   been   so    commonly   adopted   \citep[but
  see][]{Barazza2009}.  As  for bars, the influence  of environment in
the  formation and evolution  of disks  has been  widely investigated.
Physical  mechanisms   taking  place   in  galaxy  clusters,   such  as
harassment,  ram pressure  stripping, tidal  effects and  mergers, and
starvation, were  proposed for  heavily transforming the  morphology of
disks  or   even  destroying  them  \citep[][]{BoselliGavazzi2006}.
Therefore, since  the environment strongly affects  the stellar and/or
gaseous   component  of  disks,   it  is   customary  to   think  that
environmental  effects have  a direct  influence on  the  formation of
bars.

In this  Letter we derive the  bar fraction in  three different galaxy
environments ranging  from the field  to the Virgo and  Coma clusters.
The  unprecedentedly large  range of  luminosities  (or, equivalently,
stellar masses) covered by the different galaxy samples we investigate
allows us to distinguish the  effects of environment in heating galaxy
disks from those triggering bar formation.
Throughout  the paper  we assume  a flat  cosmology  with $\Omega_{\rm
  M}=0.3$, $\Omega_{\Lambda}=0.7$,  and a Hubble  parameter H$_{0}=70$
km s$^{-1}$ Mpc$^{-1}$.

\section{Galaxy sample}
\label{sec:sample}

Four galaxy samples were selected  in order to analyze three different
galaxy environments: the field, the Virgo cluster, and the Coma cluster.

The  first galaxy  sample  (hereafter field1  sample)  covers the  low
galaxy density  regime and  was taken from  \citet{Aguerri2009}.  They
selected a volume-limited sample of $\sim\,2800$ galaxies available in
the     Sloan     Digital     Sky     Survey    Data     Release     5
\citep[SDSS-DR5,][]{AdelmanMcCarthy2007}   in   the   redshift   range
$0.01\,<\,z\,<\,0.04$. The local  projected number density around each
sample galaxy  was computed  using the distance  of the galaxy  to its
fifth nearest neighbour galaxy \citep{Balogh2004}. About $80\%$ of the
galaxies    are    located    in   very    low-density    environments
($\Sigma_5\,<\,1$ Mpc$^{-2}$), whereas  the remaining $20\%$ reside in
loose      ($\Sigma_5\,\simeq\,1$     Mpc$^{-2}$)      and     compact
($\Sigma_5\,>\,10$ Mpc$^{-2}$) galaxy groups. Only the galaxies in the
lowest  local density  bin were  included in  the field1  sample, thus
containing     2389     galaxies     in    the     magnitude     range
$-24\,\lesssim\,M_r\,\lesssim\,-20$.

In order to extend our analysis to fainter field galaxies, we selected
all  the   galaxies  in  the   SDSS-DR7  \citep{Abazajian2009}  within
$2500\,<\,cz\,<\,3000$   km~s$^{-1}$  and   in  the   magnitude  range
$-21\,\lesssim\,M_r\,\lesssim\,-13$. This is  our second galaxy sample
(hereafter field2 sample) and it  consists of 352 galaxies. They are a
subsample     of     the      isolated     galaxies     studied     in
\citet{SanchezJanssen2010} and  represent a volume-limited sample
  complete out to $M_r=-15.5$.

The  third  galaxy sample  (hereafter  Virgo  sample)  is composed  of
galaxies belonging to  the Virgo cluster and was  taken from Zarattini
et  al. (in  prep.).  Cluster  members were  selected  to be  extended
sources in the SDSS-DR7  with $m_r\,<\,17.7$, within one virial radius
from  the  position  of  M87  \citep{Mamon2004},  and  with  recession
velocities available within $\pm2000$  km s$^{-1}$ with respect to the
redshift  of  Virgo cluster.  This  velocity  range  corresponds to  a
$3\sigma$  cut  on the  velocity  distribution  of  the Virgo  cluster
galaxies \citep{Binggeli1987}. For  galaxies without spectroscopy, the
cluster members were selected to have  a $g-r$ color less than 0.2 mag
above      the     value     of      the     red      sequence     fit
\citep[][]{MendezAbreu2010}.   Spurious   background  objects  were
rejected  by a  visual inspection  of their  morphology  following the
prescriptions  of  \citet{MichardAndreon2008}.   The  resulting  Virgo
sample   is  composed  of   588  galaxies   in  the   magnitude  range
$-22\,\lesssim\,M_r\,\lesssim\,-13$.

The  fourth galaxy  sample (hereafter  Coma sample)  includes galaxies
belonging    to   the    Coma   cluster    and   were    selected   by
\citet{MendezAbreu2010}.   They  are   extended  sources  in  SDSS-DR6
\citep{AdelmanMcCarthy2008} with $m_r\,<\,21$  and within a $5$-arcmin
radius from  the position of every  pointing of the  {\em Hubble Space
  Telescope\/}  Advanced  Camera  for  Surveys  ({\em  HST}-ACS)  Coma
Cluster  Treasury Survey  \citep{Carter2008}.  All  the  galaxies with
spectroscopy have  a recession  velocity within $\pm3000$  km s$^{-1}$
with respect to the Coma  cluster. This corresponds to a $3\sigma$ cut
on   the  velocity   distribution   of  the   Coma  cluster   galaxies
\citep{CollessDunn1996}.   As for  the  Virgo members,  also the  Coma
photometric members were selected to  have a $g-r$ color less than 0.2
mag above the  value of the red sequence fit  and their morphology was
inspected to  reject background objects.  The Coma  sample consists of
169  galaxies  with  $-23\,\lesssim\,M_r\,\lesssim\,-14$  and  located
mainly in the cluster center.

In an effort to homogenize the  dataset, the axial ratios, $g$ and $r$
apparent  magnitudes of  all the  galaxies  in the  four samples  were
retrieved  from the  latest  available SDSS  data  release (SDSS  III;
\citealt{Aihara2011}).
For our  analysis we considered  only the galaxies  with $b/a\,>\,0.5$
($a$ and  $b$ being  the semimajor and  semiminor axis lengths  of the
galaxies) in order to deal  with projection effects.  
It is  worth mentioning that the  galaxy images of  field2, Virgo, and
Coma samples have a  similar spatial resolution.  Indeed, the farthest
galaxies (i.e.,  those in the  Coma sample) were analyzed  by studying
the {\em HST\/} images. At the distance of the Coma cluster (100 Mpc),
the resolution of {\em HST}-ACS ($0\farcs1$) corresponds to about $50$
pc.   This gives  essentially  the same  physical  resolution as  SDSS
observations have in the field2 or in the Virgo cluster, and it allows
us to resolve bars down to sizes of $r_{\rm bar}\simeq150$ pc.
The field1  sample has a  similar resolution for the  closest galaxies
but the  mean resolution of  the sample is  $r_{\rm bar}\,\simeq\,1.3$
kpc  \citep[][]{Aguerri2009}. However, since  the field1  galaxies are
the brightest  (and, therefore, largest) galaxies in  our sample, bars
smaller than the resolution limit should be considered as nuclear bars
and therefore they are not the subject of this paper.

Another  caveat could  be  due to  the  fact that  on average  cluster
galaxies  are redder than  in the  field.  To  avoid a  possible color
bias,  we computed  the stellar  mass from  the $g-r$  color following
\citet{Zibetti2009}.

\section{Identification of disks and detection of bars}
\label{sec:detection}

We  adopted the morphological  classification of  the galaxies  in the
SSDS-DR7  spectroscopic  sample  given by  \citet{HuertasCompany2011}.
They divided the  galaxies in four morphological classes  (E, S0, Sab,
Scd) based on the automated method of \citet{HuertasCompany2008} using
learning   machines  to  analyze   the  concentration   and  asymmetry
parameters.  The main new property  of such a classification method is
that a probability of being  in each of the four morphological classes
is  associated to  each galaxy  instead of  assigning it  to  a single
class.
We classified as  a disk galaxy any galaxy with  a probability of less
than $50\%$ of  being an elliptical. Due to  the incompleteness of the
SDSS spectroscopic sample and since  several of our sample galaxies do
not have  any spectroscopic information,  we remained with  1604, 336,
228,  and 44  disk galaxies  in the  field1, field2,  Virgo,  and Coma
sample, respectively.

Following our previous  works \citep{Aguerri2009, MendezAbreu2010}, we
detected the  presence of  bars in the  different samples  by visually
inspecting  the galaxy  images. We  classified all  the  galaxies into
strong  barred, weakly  barred, and  unbarred.  The  classification of
each  sample was always  performed by  two of  us. A  caveat regarding
these criteria is that our distinction between strong and weak bars is
not  directly related  to the  contribution of  the bar  to  the total
galaxy potential, but rather refers  to a secure or possible detection
of a bar, respectively. Therefore,  the fraction of weak bars could be
understood as an estimate of the uncertainty on the bar fraction.

\begin{scriptsize}
\begin{table*}[!t]  
\begin{center}
\caption{Weighted mean and peak value of the bar fraction distributions \label{tab:statistics}}  
   \begin{scriptsize}
\begin{tabular}{c clrrr}    
\tableline  
\tableline   
\multicolumn{1}{c}{Bar Fraction Distribution} & 
\multicolumn{1}{c}{Statistical Parameter} &  
\multicolumn{1}{c}{Galaxy Property} & 
\multicolumn{1}{c}{Field Sample} &   
\multicolumn{1}{c}{Virgo Sample} & 
\multicolumn{1}{c}{Coma Sample} \\   
\multicolumn{1}{c}{(1)} & \multicolumn{1}{c}{(2)} &   
\multicolumn{1}{c}{(3)} & \multicolumn{1}{c}{(4)} &   
\multicolumn{1}{c}{(5)} & \multicolumn{1}{c}{(6)} \\   
\tableline    
                           &                        & Luminosity &     $-19.45\pm0.12$ &     $-19.50\pm0.18$ &     $-19.85\pm0.25$ \\[-1ex]
                           & \raisebox{1.5ex}{Mean} & Mass       & $9.61\pm0.06(0.12)$ & $9.94\pm0.07(0.12)$ &$10.13\pm0.09(0.13)$ \\
\raisebox{1.5ex}{Overall}  &                        & Luminosity &     $-19.07\pm0.72$ &     $-20.41\pm0.81$ &     $-20.64\pm0.45$ \\[-1ex]
                           & \raisebox{1.5ex}{Peak} & Mass       & $9.49\pm0.36(0.37)$ &$10.34\pm0.44(0.45)$ &$10.45\pm0.24(0.26)$ \\[1ex]
                           &                        & Luminosity &     $-18.81\pm0.11$ &     $-19.13\pm0.25$ &     $-19.21\pm0.29$ \\[-1ex]
                           & \raisebox{1.5ex}{Mean} & Mass       & $9.50\pm0.06(0.12)$ & $9.79\pm0.13(0.16)$ & $9.92\pm0.12(0.16)$ \\
\raisebox{1.5ex}{Ordinary} &                        & Luminosity &     $-19.07\pm0.63$ &     $-19.71\pm0.67$ &     $-20.12\pm0.78$ \\[-1ex]
                           & \raisebox{1.5ex}{Peak} & Mass       & $9.50\pm0.33(0.34)$ &$10.06\pm0.38(0.39)$ &$10.21\pm0.37(0.38)$ \\
\tableline    
\end{tabular}
\end{scriptsize}
\tablecomments{Luminosities and masses are given in $r$-band
magnitudes and $\log{({\cal M_{*}}/{\cal M}_{\sun})}$,
respectively. The errors on masses given in parentheses include the
typical uncertainty in the optical mass-to-light ratios \citep[0.1
  dex,][]{Zibetti2009}.}
\end{center}
\end{table*}
\end{scriptsize} 

\section{Bar fraction across different environments}
\label{sec:fraction}

\begin{figure*}[!ht]
 \centering
 \includegraphics[width=\textwidth]{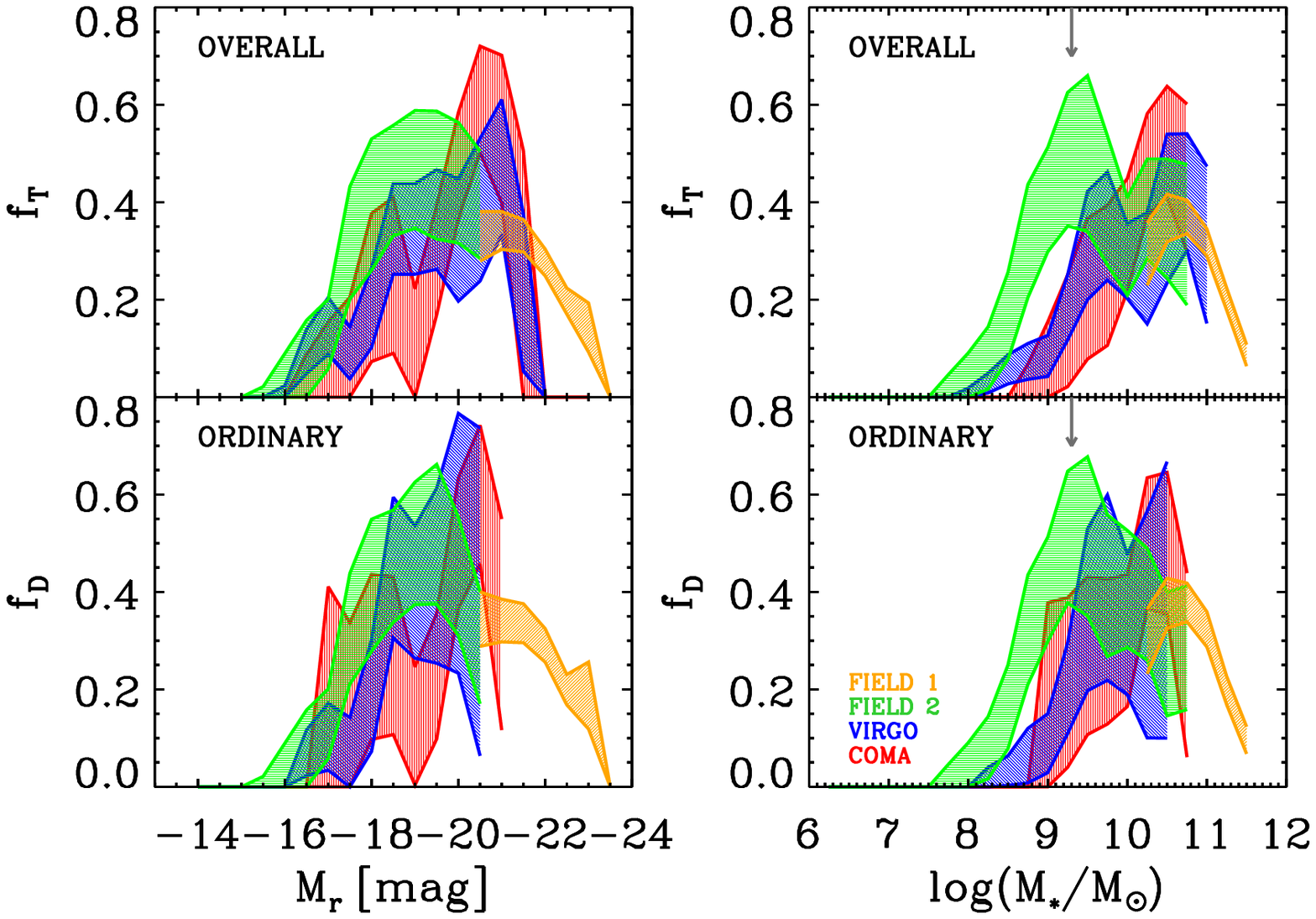}
 \caption{Bar  fraction   distribution  as  function   of  the  galaxy
   magnitudes  (left  panels)  and  masses  (right  panels).  The  bar
   fraction calculated  using all the  Hubble types ($f_{\rm  T}$) and
   only the disk  galaxies ($f_{\rm D}$) are plotted  in the upper and
   bottom panels,  respectively. The  field1, field2, Virgo,  and Coma
   samples  are showed  by the  hatched orange,  green, blue,  and red
   areas, respectively.  The grey arrow indicates  the characteristic mass
   below   which  low-mass   galaxies  start   to   be  systematically
   thicker \citep[${\cal                                     M_{*}}/{\cal
       M}_{\sun}\,\simeq\,2\,\times\,10^9$;][]{SanchezJanssen2010}.}
             \label{fig:1}%
 \end{figure*}

For each  galaxy sample we  derived the {\em ordinary\/}  bar fraction
$f_{\rm  D}$ (calculated  only for  the  disk galaxies)  and the  {\em
  overall\/} bar fraction $f_{\rm T}$ (calculated for all the galaxies
independently of their Hubble type).
Since bars  can only be triggered in  disks, $f_{\rm T}$ combines
  the luminosity distribution of  disk galaxies with their probability
  of having a bar overcoming the problem of the identification of disk
  galaxies.  This   is  always   a  major  concern   in  morphological
  classifications  dealing  with  the  measurement  of  bar  fraction.
  $f_{\rm T}$ allows  us to probe a larger  range of luminosities than
  $f_{\rm D}$.
Figure \ref{fig:1} shows  $f_{\rm D}$ and $f_{\rm T}$  as functions of
the $r$-band absolute  magnitude and mass of the  galaxies in our four
samples. To avoid issues related  to the bin size,  we applied a
  moving-average  (boxcar)  smoothing over  the  histograms using  box
  widths of  1 mag and 0.5  dex and steps of  0.5 mag and  0.25 dex in
  magnitude and mass, respectively.
The lower and upper boundaries  of the hatched areas correspond to the
bar fraction calculated by  considering only the strong (i.e., secure)
bars and both  the strong and weak (i.e.,  secure and uncertain) bars,
respectively and including their  statistical uncertainties. The
  latter  were  computed by  estimating  the  confidence intervals  on
  binomial  population  proportions  following  the  prescriptions  by
  \citet{Cameron2011}.
The values  of $f_{\rm D}$ and  $f_{\rm T}$ for the  samples of bright
(field1)  and faint  field galaxies  (field2) are  in  good agreement.
Therefore, for  studying the bar  fraction distribution we  merged the
two samples  into a  joint sample of  field galaxies  (hereafter field
sample)      with       magnitudes      in      the       range      $
-24\,\lesssim\,M_{r}\,\lesssim\,-15$     (or    $10^7\,\lesssim\,{\cal
  M_{*}}/{\cal M}_{\sun}\,\lesssim\,10^{12}$).

We calculated the weighted  mean, peak value, and corresponding errors
in  magnitude (and  mass) of  the  bar fraction  distributions of  the
samples by performing a series  of 1000 Monte Carlo simulations taking
into account the confidence intervals (Table \ref{tab:statistics}).
We  found  that  the  bar  fraction strongly  depends  on  the  galaxy
luminosity (or  mass), as previously pointed out  by \citet[][see also
  \citealt{NairAbraham2010}]{MendezAbreu2010}.
In addition, measuring $f_{\rm D}$ and $f_{\rm T}$ in different galaxy
environments allows us  to show that the bar  fraction distribution is
different  for galaxies  living  in  the field  or  in clusters.   According  to the values  given in  Table~\ref{tab:statistics}, this
  difference is  statistically significant ($>68\%$  confidence level)
  for  the  field  and  Coma  samples,  with  Virgo  sample  being  an
  intermediate case.

The  bar fraction of  field galaxies  peaks at  $M_r\,\simeq\,-19$ (or
${\cal  M_{*}}/{\cal  M}_{\sun}\,\simeq\,10^9$),  whereas the  largest
fraction   of   barred   galaxies    in   clusters   is   observed   at
$M_r\,\simeq\,-20.5$          (or          ${\cal         M_{*}}/{\cal
  M}_{\sun}\,\simeq\,10^{10.5}$).
In both of them the maximum $f_{\rm D}$ and $f_{\rm T}$ are in the range
$0.5$ to $0.6$.
Moreover, there  is marginal evidence in  our cluster samples for
two  peaks in  the overall  bar fraction  at  $M_r\,\simeq\,-20.5$ and
$-18.5$,    respectively.     This     supports    the    claims    of
\citet{NairAbraham2010} for a bimodal distribution of the bar fraction
as a function of the luminosity (or mass).

For      $M_r\,\gtrsim\,-19.5$       (or      ${\cal      M_{*}}/{\cal
  M}_{\sun}\,\lesssim\,10^{9.5}$),  the  bar  fraction  of  the  field
galaxies  is  systematically  larger  than  those of  Virgo  and  Coma
galaxies.   We speculate about  a hint  that the  bar fraction  of the
fainter galaxies in Virgo is also larger than that of Coma.
On the contrary, the overall bar fraction is larger in cluster than in
the    field     for    $-21.5\,\lesssim\,M_r\,\lesssim\,-19.5$    (or
$10^{9.5}\,\lesssim\,{\cal                                 M_{*}}/{\cal
  M}_{\sun}\,\lesssim\,10^{11}$).   The  bins  of  brighter  (or  more
massive) galaxies of the  cluster samples ($M_r\,\lesssim\,-21.5$ or $
{\cal    M_{*}}/{\cal   M}_{\sun}\,\gtrsim\,10^{11}$)    suffer   from
incompleteness.   This prevents  a  reliable comparison  for both  the
ordinary and overall bar fractions with the field galaxies.

\section{Discussion and conclusions}
\label{sec:conclusions}

In order  to study the influence  of environment on  bar formation, we
derived the fraction  of barred galaxies over a  large range of galaxy
luminosities  (or, equivalently,  stellar  masses) in  the two  nearby
benchmark clusters Virgo and Coma and  compared it to that of a sample
of field galaxies.
Both  the  distributions of  ordinary  bar  fraction  $f_{\rm D}$  and
overall bar fraction $f_{\rm T}$ were measured.
The difference of the bar fraction distributions as a function of
  galaxy luminosity (and mass) in  the field and Coma cluster is found
  to be statistically significant ($>68\%$ confidence level), with the
  Virgo cluster being an intermediate case.

Since  bars live  in disks,  these  findings allow  us to  distinguish
between the environmental processes inhibiting bar formation (heating)
or even  destroying the  host disk from  the processes  triggering the
disk instabilities which are responsible for bar formation.

We  interpret  the  decrease  of  $f_{\rm D}$  and  $f_{\rm  T}$  with
decreasing    galaxy    luminosity    observed    for    fainter    ($
M_r\,\gtrsim\,-19$  or, equivalently, less  massive galaxies  ($ {\cal
  M_{*}}/{\cal M}_{\sun}\,\lesssim\,10^{9}$) as due to the increase of
the disk  thickness. Indeed, \citet{SanchezJanssen2010}  have recently
found that the minimum of  the disk thickness distribution occurs at a
characteristic            mass           ${\cal           M_{*}}/{\cal
  M}_{\sun}\,\simeq\,2\,\times\,10^9$         (corresponding        to
$M_r\simeq\,18.5$)   below  which  low-mass   galaxies  start   to  be
systematically  thicker making  it  difficult to  develop  a bar.   We
suggest that  the values of $f_{\rm  D}$ and $f_{\rm T}$  in the field
are systematically  larger than  those in Virgo  and Coma  because the
low-mass  galaxy  disks  in   clusters  are  more  easily  heated,  or
destroyed, by  galaxy interactions and can  not develop a  bar. In the
low luminosity regime nurture and nature are acting on galaxy disks in
cluster and field, respectively.

Since the values of $f_{\rm D}$ and $f_{\rm T}$ are larger for cluster
galaxies   with   $   -21.5\,\lesssim\,M_r\,\lesssim\,-19.5$   (or   $
10^{9.5}\,\lesssim\,{\cal M_{*}}/{\cal M}_{\sun}\,\lesssim\,10^{11}$),
we  conclude  that  brighter   disks  are  strong  enough  to  survive
interactions and  form a bar.   In the high luminosity  regime nurture
and nature are acting on bar  formation in galaxy disks of cluster and
field, respectively.

From  the theoretical  point of  view, environmental  processes taking
place  in  massive  galaxy clusters  do  not  seem  to favor  such  an
enhancement  of  the  bar   population.   For  instance,  fast  galaxy
encounters  are  able to  create  bar-like  features  in galaxy  disks
depending    mainly    on    the    geometry    of    the    encounter
\citep{Mastropietro2005,  AguerriGonzalezGarcia2009}.   However, these
encounters are randomly oriented and can remove large amounts of stars
from the  galaxy outskirts  moving a galaxy  from the bright  to faint
regime  \citep{KormendyBender2012}.  Therefore,  we  suggest that  the
enhanced bar  fraction observed in Virgo  and Coma is not  only due to
processes taking place in clusters.

One   possible  explanation   could  be   related  to   the  different
morphological  mixing in the  field and  clusters.  In  the luminosity
range  we probe,  the galaxy  population in  clusters is  dominated by
lenticular       galaxies       \citep{Binggeli1988}.        Recently,
\citet{Cameron2010},   \citet{Masters2011},  and  \citet{Marinova2012}
have  found  that  the  fraction  of barred  galaxies  is  higher  for
lenticulars  than for  spirals. This  is  contrary to  the results  of
\citet{Laurikainen2007}  and  \citet{Aguerri2009}   and  also  to  the
structural properties of lenticular galaxies (i.e., large bulge, thick
disk, and little gas) which should inhibit the formation of bars.
However, even  if the higher  fraction of barred galaxies  in clusters
was  confirmed to be  due to  lenticulars, this  would point  again to
environmental processes  acting in triggering bar  formation since the
typical   environmental  mechanisms   observed   in  clusters   (i.e.,
harassment, high-speed  encounters, or ram-pressure  stripping) should
not  be able  to make  disk galaxies,  in general,  more prone  to bar
instabilities.

On the other hand, slow galaxy  encounters have been suggested to be a
very   efficient  mechanism   in  inducing   the  formation   of  bars
\citep{Noguchi1988}. These encounters are not likely to happen in rich
clusters,  like Virgo and  Coma, but  they are  more common  in galaxy
groups.  Therefore,  we tentatively  suggest that bar  formation could
preferentially occur in small-size galaxy groups before they fall into
the cluster.  If this is  the case, we  expect to observe a  large bar
fraction in galaxy groups. Moreover, some of the bars we are observing
in clusters should represent the population of genuine bars induced by
interactions  and  should have  different  observable properties  with
respect to spontaneous bars as predicted by \citet{MiwaNoguchi1998}.

To summarize,  we suggest a bimodality  in the role  of environment on
bars, since interactions  can trigger or inhibit the  bar formation in
galaxy  disks depending  on their  luminosity. The  disks  of brighter
galaxies are  strong enough both  to maintain their cold  structure and
survive close  interactions, which trigger the bar  formation when the
galaxies are  probably in  a pre-cluster stage.  On the  contrary, the
disks of fainter galaxies are more fragile and hot. They can be heated
or even  destroyed by galaxy interactions which  inhibit the formation
of bars.
Classification methods  (e.g., visual inspection,  concentration cuts,
color-based cuts) used to select a disk galaxy sample are known to add
biases.  These  are usually  considered  as  the  main source  of  the
different  findings   about  the  influence  of   environment  in  bar
formation.
Our results highlight the importance of studying galaxy samples
which have been carefully selected in luminosity to avoid biases when
dealing with bar statistics. We argue that most of the controversial
results about the relationship between environment and bar fraction
could be explained in terms of the different luminosity ranges covered
by the galaxy samples studied so far.

\acknowledgments

We  thank  V.  Debattista,  A.   de  Lorenzo-Caceres,  I.  Perez,  and
P. Sanchez-Blazquez  for useful discussions and  suggestions.  We also
thank the referee for constructive comments.  J.M.A. and J.A.L.A.  are
partially    funded   by    the   Consolider-Ingenio    2010   Program
(CSD2006-00070)  and Collaboration  ESTALLIDOS (AYA2010-21887-C04-04).
E.M.C.   is supported  by  Padua University  grants 60A02-5934/09  and
60A02-1283/10.

\end{document}